# Developing an Activity-Based Costing Approach to Maximize the Efficiency of Customer Relationship Management Projects

**Mahmood Shafiee[1], Golriz Amooee[2], Yaghoub Farjami[3]**

[1] **Department of Industrial Engineering, Faculty of Engineering**
**Tarbiat Modares University, Tehran, Iran**
*E-mail:* **mshafiee@modares.ac.ir**

[2] **Corresponding Author- Department of Information Technology, University of Qom**
**Qom, Iran**
*E-mail:* **Golriz.Amooee@gmail.com**

[3] **Department of Information Technology, University of Qom, Qom, Iran**
*E-mail:* **farjami@gmail.com**

**Abstract**

In today's competitive environment, profitability analysis is not just about looking at the profit and loss statement. It is more about knowing which of your customers are making you money and which are losing you money. This paper considers how activity-based costing approach may complement a customer relationship management effort. The model presented in this paper combines the principles of activity-based costing with performance measurement. Applying this model helps managers understand the true costs of providing products and services, and the factors that drive these costs, while addressing other concerns such as customer satisfaction. This approach has the potential to integrate all business processes around the requirements of significant profitable customers, a fact that most of the previous researches fail to acknowledge.

## 1. Introduction

Organizations have increasingly recognized the importance of managing customer relationships, and many organizations are turning to customer relationship management (CRM) to better serve customers and facilitate closer relationships with them [1]. At its core, customer relationship management is about acquiring customers, knowing them well, providing services and anticipating their needs [2].

Customers differ in their costs to serve. Some customers tend to be considerably more costly to serve than others. Despite these differences, marketing scholars have not been very sensitive to the issue of differential customer costs. Even if a marketer was sensitive to differences in cost to serve, accounting systems were not capable of tracking the cost to serve of individual customers [3]. So accountants simply allocated the cost of augmented services evenly across to all customers. But in the recent years, the widespread acceptance of activity-based costing has allowed firms to precisely allocate overhead costs to specific customers [4].

The precise measurement of customer cost has opened many firms` eyes about the importance of cost-to-serve in guiding customer management strategies. For this purpose, activity based costing (ABC) was designed and becomes a tool for determining true customer costs and will provides managers with insight into customer profitability. Despite the advances activity based costing system offers, there is little research on how the customer cost information affects firm strategies.

Our goal in this paper is to address this gap in the literature by combining activity-based costing with customer relationship management. The paper focuses on what actions a firm should take given the additional information obtained from first period purchases on customer revenues and cost. And how should a firm serve its customers in order to dynamically improve its profitability using the mixture of high and low profitable customers? With the help of the proposed model, all firms can set strategies based on customers` satisfaction and their cost information.

The outline of the paper is as follows: we present an overview on the traditional costing systems and the need for new methods. We also address the issues related to successful customer relationship management implementation and suggest activity-based costing to maximize its benefits. Finally, we propose a new framework to categorize customers. We use TOPSIS to prioritize these customers and maximize the efficiency of customer relationship management projects by proper customer profitability analysis.

## 2. Literature Review

Traditional costing systems misleading financial reports by using only some of the cost drivers and provided distorted information. Progression of production technologies and several other factors has changed product cost structure greatly and increased overhead costs and cause a sharp reduction in direct labor and material costs instead. In such areas, organizations increasingly seek to improve their costing systems. Relevant cost information plays an important role in management decisions. Providing such information needs new methods to provide management required information. For this purpose, activity- based costing has received considerable attention in the academic researches.

Evans and Bellamy [5] argue the necessity of developing this new method in order to cost the services of Public Sector for a better management. In the Macedonia University of Thessaloniki Greece, Vazakidis and Karagiannis [6] for the first time presented a model of cost accounting for the Department of Applied Informatics of University of Macedonia in Thessalonica, more for internal information. Finally in 2008, they applied a new model of Activity-Based Costing and Activity-Based Management in a tourist organization [7], so as to point out the usefulness of the method as a tool and source of information for the administration.

Narayanan [8] investigates the benefits of activity based pricing compared to traditional pricing models using a static model, when the monopolist is able to price based on the metered use of services in a B2B environment. He concludes that activity based pricing is beneficial when there is more variability in the cost-to serve among customers in a monopoly setting.

Haenlein and Kaplan [9] address the consequence of cost based pricing strategy on a firm's long-term profitability, when firms are vulnerable to the negative word of mouth which cost based pricing may generate.

Finally, some researches are related to customer relationship management which emphasizes the importance of identifying the right customers for a successful customer relationship management program [10] [11]. Researchers have, therefore, focused on the identification of good customers [12] by estimating customer lifetime value [13] [14] and providing them with differentiated value propositions through different price levels [15]. However, none of these papers addresses the help of activity-based costing in an effective customer relationship management.

## 3. Customer Profitability Analysis

Customer profitability analysis was conducted across one financial year for all of the company`s customers which demand for any of company`s activities or products during that period. According to the results calculated with the ABC model, customer profitability varied greatly.

Here a question should be raised. What was the main cause of these discrepancies in the generated profits by different customers? How can we prioritize our customers to increase profitability? To answer these questions first the drivers of customer profitability should be identified.

2.1 Customers Profiles

Previous analytical models worked under this premise that high volume customers are our profitable customers. Many companies think that improving their customer services to expand their market share, will create value and loyalty among these customers and higher profits will be generated. But studies on customer profitability revealed that high volume customers are not necessary profitable. Different customers demand different combination of company`s activities so customer profitability analysis must work through all customer related activities.

The drivers of customer profitability which we used, are based on ABC/CRM model proposed in [16] (see figure 1). This model shows how different customers, individually or as a group, contribute to profitability. As a result the information that the model provides can help the company to determine which customers are the most profitable, what efforts should be made toward customer related improvements and whether processes are customer value added or not.

The main drivers are as follows:

- **Value "from" customer:** The value each customer produced for the firm has intuitive appeal as a marketing concept, because in theory it represents exactly how much each customer is worth in monetary terms, and therefore exactly how much a marketing department should be willing to spend to acquire each customer. Since all customers are not financially attractive, it is critical for companies to measure the customers' level of profitability according to their lifetime value. Gupta et al proposed a calculation model as shown in Eq. (1):

$$\text{Customer lifetime value} = \sum_{t=0}^{T} \frac{(p_t - c_t)r_t}{(1+i)^t} - AC \quad (1)$$

Where
$P_t$ = price paid by a consumer at time t;
$C_t$ = direct cost of servicing the customer at time t;
i = discount rate or cost of capital for the firm;
$r_t$ = probability of customer repeat buying or being "alive" at time t;
AC = acquisition cost;
T = time horizon for estimating CLV.

- **Value "to" customer:** Although value from customers is important, but real value to companies lies in the value they create for their customers, which is given back to them accordingly by those same customers. This parameter explicitly incorporates the possibility that a customer may defect to competitors in the future. This indicator is derived largely from the quality and reliability of your products and services. Value can be defined as the perceived benefits compared with the perceived costs. However, it is not as simple as that because value lies in a customer's mind. What is value for one customer is not necessarily value for another one.

One way to calculate value to customer i, is as shown in Eq. (2):

$$V2C = \sum_{j=1}^{n} \alpha_{ij} \frac{(B_{ij} - C_{ij})}{C_{ij}} \quad 0 \leq \alpha_{ij} \leq 1, j = 1, \ldots, n \quad (2)$$

Where
$\alpha_{ij}$ is the importance of each parameter in customer's viewpoint;
$B_{ij}$ is the benefits of parameter j for customer i;
$C_{ij}$ is the costs of parameter j for customer i.

Many features are contributed to form this value. All customers weigh up the perceived benefits of a purchase against the perceived costs differently. What they accrue to the benefits of a product changes with their personality, their experiences and the environment. What they perceive as a cost is also different from one person to another person. Creating value for customers in specified items will lead to customer's loyalty and satisfaction, and results in increased business and therefore further profitability. To estimate customer value we can use a questionnaire. We cover all customer`s key requirements under these categories: Planning and financial resources, Understanding and friendliness, Control and fairness, Options and alternatives, Information and communication. We design a questionnaire with total of 50 questions and ask customers to fill it. The total score is obtained by adding up the scores customers give to each provided service. With this number we can apprize the value created for each customer.

We use the discussed drivers and create the following customer groups:

1) Passenger: This condition only occurs in those circumstances which there are a significant sudden need. This is often because there was no other option for the customer, besides the referred organization. These customers may also settle for less expensive products even with lower quality, they will be also looking for cheaper alternatives.

2) Cost to serve: These customers buy any time they want. Customer perceived value (CPV) is high but there is no value created for the organization. These customers should be treated with care. they overuse firm`s resources but there is no significant benefit for the organization. So the firm should always be careful not to allocate too many resources for these customers.

3) Challenger: In this case the firm must have the ability to quickly identify the sales, marketing or customer management issues involved and propose practical solutions, otherwise it will lose customers to competitors. If we don`t solve customer problems, sooner or later, they will reduce the spending or completely go for low cost substitutes. They are more prudent and look out for more options.

4) Noteworthy: The firm should make this group brand loyal and privilege them by opportunity development and more options. An increase in their lifetime value will lead to firm's profitability. Therefore, uses of effective customer management and aim to retain customers and grow major business with them should be developed. This group will always look for sales promotions. The business case firmly grounded in a current and objective knowledge of customer needs, appropriate systems and technology platforms and clear, measurable deliverables.

To evaluate the proposed model, we used customers' last year information of a manufacturing company. According to the different multi-criteria decision making methods in similar projects and with regard to our study characteristics, we used TOPSIS (Technique for Order Preference by Similarity to Ideal Solution) and SAW (Simple Additive Weighting) methods as a quantitative approach to prioritize and analyze these customers. Each of these methods is presented in later sections.

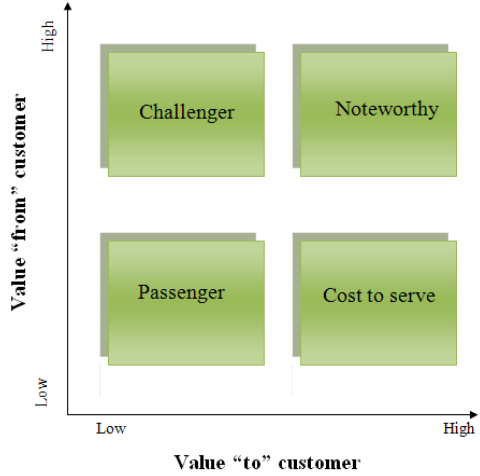

Figure 1 Customers' profiles in term of their value

## 4. Analyzing and Prioritizing Customers based on SAW

Simple Additive Weighting (SAW) is the easiest Multi-criteria decision making method. This method was proposed by Hwang and Yoon in 1981 [17]. To deal with the customer ranking problem we used the two discussed criteria. SAW method can be outlined as following steps:

**Step 1:** Construct decision matrix. For this purpose we should quantify qualitative data. Likert scale [18] is the most widely used approach to scaling responses in survey researches. This approach emerges from collective responses to a set of items, and the format in which responses are scored along a range. The format of a typical five-level Likert item, for example, is shown in table 1.

Table 1: Likert scale using five-level Likert item

| Qualitative | Strongly disagree | Disagree | Neither agree nor disagree | Agree | Strongly agree |
|---|---|---|---|---|---|
| Quantitative | 1 | 3 | 5 | 7 | 9 |

**Step 2:** Compute the normalized decision matrix. The normalized value $r_{ij}$ is calculated as in Eq. (3):

$$Rij = \frac{aij}{\sqrt{(\sum_{i=1}^{m} aij^2)}} \quad (3)$$

We also can normalized values based on Eq(4) and Eq(5) Where R* is associated with advantage criteria, and R⁻ is associated with cost criteria:

$$R^* = \frac{R_i - R_{min}}{R_{max} - R_{min}} \quad (4)$$

$$R^- = \frac{R_{max} - R_i}{R_{max} - R_{min}} \quad (5)$$

Since both of our values are advantage criteria, only Eq(4) has been used. You can see the results in appendix A.

**Step 3:** Calculate the weighted normalized decision matrix. The weighted normalized value $v_{ij}$ is calculated as Eq. (6):

$$V_{ij} = w_i * r_{ij} \quad (6)$$

Where $w_i$ is the weight of the ith attribute or criterion, and:

$$\sum_{i=1}^{n} w = 1 \quad (7)$$

You can see the results in Appendix C.
We used Shannon's method [19] to calculate the weights through the following steps:

(1) Normalize the evaluation index as in Eq(8):

$$Pij = \frac{Xij}{\sum_{j} Xij} \quad (8)$$

(2) Calculate entropy measure of every index as shown in Eq(9):

$$ej = -(\ln(m))^{-1} \sum_{j=1}^{n} Pij \ln(Pij) \quad (9)$$

(3) Define the divergence through Eq(10):

$$DIV_g = 1 - e_j \quad (10)$$

The more the divj is, shows the importance of the criterion jth.

(4) Obtain the normalized weights of indexes as shown in Eq(11):

$$wj = \frac{DIVj}{\sum_{j} DIVj} \quad (11)$$

Weights obtained for "value to the firm" and "value to the customer" are 0.492 and 0.508 respectively. You can see the calculation in Appendix B.

**Step 4:** Selecting the best choice. The best choice is obtained from the following Eq. (12):

$$A^* = \{A_i \mid \max_i \sum_{j=1}^{m} w_j r_{ij} \} \quad (12)$$

Finally, the choices are ranked based on descending order of $A^*$. Results are shown in table 2.

Table 2: Final results of SAW

| Customers | $A^*$ | Rank | Customers | $A^*$ | Rank |
|---|---|---|---|---|---|
| C1 | 0.679 | 5 | C25 | 0.120 | 42 |
| C2 | 0.421 | 23 | C26 | 0.317 | 29 |
| C3 | 0.809 | 3 | C27 | 0.502 | 13 |
| C4 | 0.131 | 41 | C28 | 0.436 | 21 |
| C5 | 0.062 | 46 | C29 | 0.111 | 43 |
| C6 | 0.422 | 22 | C30 | 0.702 | 4 |
| C7 | 0.232 | 35 | C31 | 0.497 | 16 |
| C8 | 0.100 | 44 | C32 | 0.153 | 39 |
| C9 | 0.064 | 45 | C33 | 0.508 | 11 |
| C10 | 0.385 | 25 | C34 | 0.362 | 26 |
| C11 | 0.280 | 33 | C35 | 0.501 | 14 |
| C12 | 0.464 | 19 | C36 | 0.045 | 47 |
| C13 | 0.506 | 12 | C37 | 0.310 | 30 |
| C14 | 0.210 | 36 | C38 | 0.197 | 37 |
| C15 | 0.325 | 28 | C39 | 0.922 | 1 |
| C16 | 0.331 | 27 | C40 | 0.509 | 10 |
| C17 | 0.259 | 34 | C41 | 0.588 | 8 |
| C18 | 0.386 | 24 | C42 | 0.499 | 15 |
| C19 | 0.136 | 40 | C43 | 0.660 | 6 |
| C20 | 0.483 | 17 | C44 | 0.468 | 18 |
| C21 | 0.292 | 32 | C45 | 0.438 | 20 |
| C22 | 0.310 | 31 | C46 | 0.821 | 2 |
| C23 | 0.601 | 7 | C47 | 0.513 | 9 |
| C24 | 0.187 | 38 | | | |

\* $C_i$ indicates our customers.

## 5. Analyzing and Prioritizing Customers based on TOPSIS

TOPSIS (Technique for Order Preference by Similarity to Ideal Solution), is proposed by Hwang and Yoon (1981). The basic principle of TOPSIS is that, chosen alternatives should have the shortest distance from the ideal solution and the farthest distance from the negative-ideal solution. According to [20], some advantages of TOPSIS are as follows:

- A sound logic that embodies the rational of human choice.
- A simple computation process that can be easily programmed into a spreadsheet.
- A scalar value that accounts for both the best and worst alternative at the same time.

To deal with the customer ranking problem we used the two discussed criteria for TOPSIS method. According to [21], TOPSIS can be outlined as following steps:

**Step 1 – 3: Exactly like SAW method.**

**Step 4:** Determine the ideal and negative-ideal solutions based on Eq(13) and Eq(14).

$$A^* = \{(Max v_{ij} \mid j^*), (\min v_{ij} \mid j^-)\} \quad (13)$$

$$A^- = \{(Min v_{ij} \mid j^*), (Max v_{ij} \mid j^-)\} \quad (14)$$

Where j* is associated with advantage criteria, and j⁻ is associated with cost criteria.

**Step 5:** Calculate the separation measures, using the n-dimensional Euclidean distance. The separation of each alternative from the ideal solution is given as Eq(15):

$$d_i^* = \left\{ \sum_{j=1}^{n} (V_{ij} - V_j^*)^2 \right\}^{\frac{1}{2}}, \quad (i = 1, 2, ..., m) \quad (15)$$

Similarly, the separation from the negative-ideal solution is given as Eq(16):

$$d_i^- = \left\{ \sum_{j=1}^{n} (V_{ij} - V_j^-)^2 \right\}^{\frac{1}{2}}, \quad (i = 1, 2, ..., m) \quad (16)$$

**Step 6:** Calculate the relative closeness to the ideal solution. The relative closeness of the alternative aj with respect to $A^*$ is defined as Eq(17):

$$cli^* = \frac{di^-}{di^- + di^*} \quad (17)$$

You can see the results in Appendix D.

**Step 7:** Rank the preference order. According to the closeness coefficient, we can understand the assessment status of each alternative and determine the ranking order of them.

| Customers | Cli | Rank | Customers | Cli | Rank |
|---|---|---|---|---|---|
| C1 | 0.611 | 6 | C25 | 0.084 | 42 |
| C2 | 0.370 | 23 | C26 | 0.284 | 28 |
| C3 | 0.792 | 3 | C27 | 0.462 | 11 |
| C4 | 0.088 | 41 | C28 | 0.396 | 20 |
| C5 | 0.024 | 46 | C29 | 0.068 | 44 |
| C6 | 0.377 | 22 | C30 | 0.696 | 4 |
| C7 | 0.189 | 35 | C31 | 0.426 | 17 |
| C8 | 0.078 | 43 | C32 | 0.113 | 40 |
| C9 | 0.027 | 45 | C33 | 0.495 | 10 |
| C10 | 0.336 | 26 | C34 | 0.338 | 25 |
| C11 | 0.240 | 33 | C35 | 0.432 | 15 |
| C12 | 0.433 | 13 | C36 | 0.000 | 47 |
| C13 | 0.433 | 14 | C37 | 0.266 | 32 |
| C14 | 0.169 | 36 | C38 | 0.154 | 37 |
| C15 | 0.282 | 29 | C39 | 0.893 | 1 |
| C16 | 0.292 | 27 | C40 | 0.517 | 9 |
| C17 | 0.216 | 34 | C41 | 0.535 | 8 |
| C18 | 0.341 | 24 | C42 | 0.430 | 16 |
| C19 | 0.122 | 39 | C43 | 0.663 | 5 |
| C20 | 0.421 | 18 | C44 | 0.412 | 19 |
| C21 | 0.271 | 30 | C45 | 0.383 | 21 |
| C22 | 0.270 | 31 | C46 | 0.808 | 2 |
| C23 | 0.595 | 7 | C47 | 0.448 | 12 |
| C24 | 0.148 | 38 | | | |

Table 3: Final results of TOPSIS

* $C_i$ indicates our customers.

## 6. Prioritizing Strategy

Based on different methods which we used to prioritize customers (SAW and TOPSIS) in this paper, and due to different rankings obtained for each of them, we used an integration method (Copeland) to resolve the conflicts between these ranks. Copeland's method or Copeland's pairwise aggregation method is a condorcet method in which candidates are ordered by the number of pairwise victories, minus the number of pairwise defeats [22]. For example, for these four customers we have:

Table 4: Ranks of a group of customers in our two method

| Customer | Rank in TOPSIS | Rank in SAW |
|---|---|---|
| C10 | 26 | 25 |
| C11 | 33 | 33 |
| C12 | 13 | 19 |
| C13 | 14 | 12 |

Table5 : Majority rule for our sample customers

|  | C10 | C11 | C12 | C13 | $\sum C$ |
|---|---|---|---|---|---|
| C10 | - | M | X | X | 1 |
| C11 | X | - | X | X | 0 |
| C12 | M | M | - | X | 2 |
| C13 | M | M | X | - | 2 |
| $\sum R$ | 2 | 3 | 0 | 0 | |

Copeland score of each customer is calculated as Eq(18):
$C_i$ score = victories - defeats
$C_i$ score = $\sum C - \sum R$                      (18)

$C_{10}=1-2=-1$
$C_{11}=0-3=-3$
$C_{12}=2-0=2$
$C_{13}=2-0=2$

Therefore the customers' ranks are as follows:
$C_{13}=C_{12}>C_{10}>C_{11}$

The final ranking, for all customers, is as follows:

$C39 > C46 > C3 > C30 > C1 = C43 > C23 > C41 > C40 > C47 > C33 > C27 > C13 > C35$
$> C42 > C31 = C12 > C44 > C20 > C45 > C28 = C6 > C2 = C18 > C10$
$> C34 > C16 > C15 = C26 > C37 > C21 = C22 > C11 > C17 > C7 > C14$
$> C38 > C24 > C19 = C32 > C4 > C25 > C8 > C29 > C9 > C5 > C36$

Based on the priorities obtained for each customer from the model, organizations must first allocated their resources to customers with higher priority and then move toward lower-priority customers.

## 7. Conclusions

One of the main components of customer relationship management is the ability to measure the profitability of the customer. Customer profitability analysis is a new business approach that reflects required strategies for profitability growth. Success in this analysis depends on the accuracy of data. The cost data provided by activity based costing systems, allows for more accurate determination of customer profitability. This paper has tried to use this new cost system, to develop a new model to maximize the efficiency of customer relationship management projects. By using the profiles obtained from the model, appropriate strategies to retain and maintain profitable customers can be adopted. The results show that the company`s profit was after 29.8 percent (14 out of 47) of its customers. At this stage the profit was 88 percent of the actual profit. The remaining customers were either broke even or created losses. The research also shows that even an unprofitable customer can be worthwhile, because it is usually much easier to improve an existing unprofitable customer into a profitable one than to find a new profitable customer and it also will cost less.

**Appendix A**
The normalized decision matrix

$$\begin{pmatrix}
0.907 & 0.521 \\
0.641 & 0.268 \\
0.854 & 0.778 \\
0.209 & 0.078 \\
0.111 & 0.028 \\
0.899 & 0.092 \\
0.372 & 0.136 \\
0.111 & 0.093 \\
0.111 & 0.032 \\
0.622 & 0.221 \\
0.489 & 0.136 \\
0.529 & 0.419 \\
0.949 & 0.200 \\
0.289 & 0.156 \\
0.555 & 0.166 \\
0.422 & 0.268 \\
0.422 & 0.147 \\
0.720 & 0.156 \\
0.160 & 0.119 \\
0.809 & 0.259 \\
0.289 & 0.294 \\
0.555 & 0.140 \\
0.555 & 0.633 \\
0.244 & 0.147 \\
0.160 & 0.092 \\
0.622 & 0.107 \\
0.622 & 0.419 \\
0.555 & 0.354 \\
0.209 & 0.044 \\
0.672 & 0.722 \\
1 & 0.151 \\
0.209 & 0.114 \\
0.489 & 0.521 \\
0.372 & 0.354 \\
0.880 & 0.239 \\
0.111 & 0 \\
0.489 & 0.186 \\
0.289 & 0.133 \\
0.808 & 1 \\
0.372 & 0.604 \\
0.773 & 0.460 \\
0.889 & 0.230 \\
0.489 & 0.778 \\
0.720 & 0.294 \\
0.800 & 0.188 \\
0.672 & 0.924 \\
0.808 & 0.309
\end{pmatrix}$$

**Appendix B**
Shannon's method [19] for calculating weights

| Pij ( A ) | Pij ( B ) | |
|---|---|---|
| 0.036523 | 0.038029 | |
| 0.025811 | 0.019577 | |
| 0.034388 | 0.05671 | |
| 0.008416 | 0.005663 | |
| 0.00447 | 0.002024 | |
| 0.0362 | 0.006735 | |
| 0.014979 | 0.009902 | |
| 0.00447 | 0.006783 | |
| 0.00447 | 0.002352 | |
| 0.025046 | 0.016111 | |
| 0.019691 | 0.009937 | |
| 0.021301 | 0.030564 | |
| 0.038214 | 0.01459 | |
| 0.011637 | 0.011347 | |
| 0.022348 | 0.012096 | |
| 0.016993 | 0.019577 | |
| 0.016993 | 0.010702 | |
| 0.00447 | 0.011368 | |
| 0.006443 | 0.008688 | |
| 0.032576 | 0.018866 | $e_j$ (A) = 0.9446 |
| 0.011637 | 0.02146 | $e_j$(B) = 0.9198 |
| 0.022348 | 0.01022 | |
| 0.022348 | 0.046156 | |
| 0.009825 | 0.010756 | DIV (A) =0.0553 |
| 0.006443 | 0.006735 | DIV (B) = 0.0802 |
| 0.025046 | 0.007807 | |
| 0.025046 | 0.030564 | |
| 0.022348 | 0.025851 | W (A) = 0.4083 |
| 0.008416 | 0.003186 | W (B) = 0.5917 |
| 0.02706 | 0.052662 | |
| 0.040267 | 0.010985 | |
| 0.008416 | 0.008343 | |
| 0.019691 | 0.038029 | |
| 0.014979 | 0.025851 | |
| 0.035435 | 0.017433 | |
| 0.00447 | 0 | |
| 0.019691 | 0.013592 | |
| 0.011637 | 0.009681 | |
| 0.032536 | 0.072924 | |
| 0.014979 | 0.044016 | |
| 0.031127 | 0.033538 | |
| 0.035798 | 0.016795 | |
| 0.019691 | 0.05671 | |
| 0.028993 | 0.02146 | |
| 0.032214 | 0.0137 | |
| 0.02706 | 0.067391 | |
| 0.032536 | 0.022533 | |

**Appendix C**
The weighted normalized decision matrix

| Customer | V2C | V2F |
|---|---|---|
| C1 | 0.370318 | 0.308568 |
| C2 | 0.261713 | 0.158851 |
| C3 | 0.348679 | 0.460147 |
| C4 | 0.085332 | 0.045948 |
| C5 | 0.04532 | 0.016419 |
| C6 | 0.367052 | 0.054648 |
| C7 | 0.151884 | 0.080348 |
| C8 | 0.04532 | 0.055037 |
| C9 | 0.04532 | 0.019087 |
| C10 | 0.253956 | 0.13073 |
| C11 | 0.199653 | 0.080626 |
| C12 | 0.215985 | 0.248003 |
| C13 | 0.387466 | 0.118383 |
| C14 | 0.117996 | 0.092067 |
| C15 | 0.2266 | 0.098149 |
| C16 | 0.172298 | 0.158851 |
| C17 | 0.172298 | 0.086839 |
| C18 | 0.293968 | 0.092243 |
| C19 | 0.065326 | 0.070493 |
| C20 | 0.330306 | 0.153081 |
| C21 | 0.117996 | 0.17413 |
| C22 | 0.2266 | 0.082924 |
| C23 | 0.2266 | 0.374515 |
| C24 | 0.099623 | 0.087274 |
| C25 | 0.065326 | 0.054648 |
| C26 | 0.253956 | 0.063348 |
| C27 | 0.253956 | 0.248003 |
| C28 | 0.2266 | 0.209758 |
| C29 | 0.085332 | 0.02585 |
| C30 | 0.27437 | 0.427301 |
| **C31** | **0.408289** | 0.089134 |
| C32 | 0.085332 | 0.067699 |
| C33 | 0.199653 | 0.308568 |
| C34 | 0.151884 | 0.209758 |
| C35 | 0.359294 | 0.14145 |
| **C36** | **0.04532** | **0** |
| C37 | 0.199653 | 0.110289 |
| C38 | 0.117996 | 0.078549 |
| C39 | 0.329898 | **0.591711** |
| C40 | 0.151884 | 0.357146 |
| C41 | 0.315607 | 0.272129 |
| C42 | 0.362969 | 0.136278 |
| C43 | 0.199653 | 0.460147 |
| C44 | 0.293968 | 0.17413 |
| C45 | 0.326631 | 0.111165 |
| C46 | 0.27437 | 0.54682 |
| C47 | 0.329898 | 0.182838 |

**Appendix D**
Separations from the ideal/negative-ideal solution

| Customer | Si * | Si - | Ci |
|---|---|---|---|
| C1 | 0.240002 | 0.479002 | 0.666202 |
| C2 | 0.403353 | 0.299989 | 0.426519 |
| C3 | 0.132139 | 0.537441 | 0.802654 |
| C4 | 0.606087 | 0.062762 | 0.093836 |
| C5 | 0.657827 | 0.013645 | 0.020321 |
| C6 | 0.44926 | 0.403082 | 0.472911 |
| C7 | 0.531483 | 0.148515 | 0.218405 |
| C8 | 0.634886 | 0.045737 | 0.067199 |
| C9 | 0.656217 | 0.015862 | 0.023601 |
| C10 | 0.428569 | 0.281531 | 0.396467 |
| C11 | 0.497847 | 0.203474 | 0.290129 |
| C12 | 0.372688 | 0.295997 | 0.442656 |
| C13 | 0.394205 | 0.437143 | 0.525824 |
| C14 | 0.550457 | 0.118487 | 0.177125 |
| C15 | 0.468395 | 0.239959 | 0.338755 |
| C16 | 0.464441 | 0.205945 | 0.307204 |
| C17 | 0.512193 | 0.173766 | 0.253318 |
| C18 | 0.438795 | 0.318886 | 0.420871 |
| C19 | 0.443896 | 0.063656 | 0.125418 |
| C20 | 0.377222 | 0.37689 | 0.49978 |
| C21 | 0.501019 | 0.170662 | 0.254082 |
| C22 | 0.479514 | 0.235958 | 0.329794 |
| C23 | 0.289372 | 0.38444 | 0.570545 |
| C24 | 0.568666 | 0.099146 | 0.148464 |
| C25 | 0.617642 | 0.051795 | 0.077372 |
| C26 | 0.479281 | 0.265006 | 0.356054 |
| C27 | 0.344236 | 0.331562 | 0.490623 |
| C28 | 0.389756 | 0.285155 | 0.422507 |
| C29 | 0.618685 | 0.054245 | 0.08061 |
| C30 | 0.215547 | 0.455413 | 0.678748 |
| C31 | 0.417659 | 0.457881 | 0.52297 |
| C32 | 0.592682 | 0.075141 | 0.112517 |
| C33 | 0.350462 | 0.32042 | 0.47761 |
| C34 | 0.450152 | 0.219053 | 0.327333 |
| C35 | 0.379121 | 0.408152 | 0.518438 |
| C36 | 0.667809 | 0 | 0 |
| C37 | 0.47699 | 0.212868 | 0.308567 |
| C38 | 0.55898 | 0.111563 | 0.166377 |
| C39 | 0.097587 | 0.606055 | 0.861311 |
| C40 | 0.374008 | 0.325098 | 0.46502 |
| C41 | 0.289563 | 0.40541 | 0.583347 |
| C42 | 0.382662 | 0.41133 | 0.518053 |
| C43 | 0.281799 | 0.427949 | 0.602959 |
| C44 | 0.375072 | 0.34169 | 0.476713 |
| C45 | 0.412085 | 0.362177 | 0.467771 |
| C46 | 0.170835 | 0.536476 | 0.758473 |
| C47 | 0.353524 | 0.385473 | 0.521617 |

**Dr. Mahmood Shafiee** received his MS degree in Industrial Engineering from Sharif University of Technology, Tehran, in 2006 and PhD degree (with a first class Honors) in Industrial Engineering from Iran University of Science and Technology, Tehran, in 2010. He has more than 40 papers published in international journals and conference proceedings. His papers have been appeared in *Reliability Engineering and System Safety, IIE Transactions, Journal of Risk and Reliability, Asia-Pacific Journal of Operational Research, International Journal of Advanced Manufacturing Technology, Applied Stochastic Models in Business and Industry, Communications in Statistics-Theory and Methods, International Journal of Quality and Reliability Management,* etc.

**Golriz Amooee** was born in Tehran, Iran in 1987. She received her B.S. degree in Information Technology from Islamic Azad University, Parand Branch, Iran, in 2009 and currently, she is a M.S. student in the Department of Information Technology at University of Qom, Iran. She specializes in the field of Customer Relationship Management (CRM), Information Security Management and ISO 27001.